# A preliminary XML-based search system for planetary data

F. Carraro[1*], S. Fonte[1], D. Turrini[1], M. C. De Sanctis[2], L. Giacomini[1]

[1] INAF-IFSI, Via Fosso del Cavaliere 100, 00133, Rome, Italy
[2] INAF-IASF, Via Fosso del Cavaliere 100, 00133, Rome, Italy

* francesco.carraro@ifsi-roma.inaf.it

## Abstract

Planetary sciences can benefit from several different sources of information, i.e. ground-based or near Earth-based observations, space missions and laboratory experiments. The data collected from these sources, however, are spread over a number of smaller, separate communities and stored through different facilities: this makes it difficult to integrate them. The IDIS initiative, born in the context of the Europlanet project, performed a pilot study of the viability and the issues to be overcome in order to create an integrated search system for planetary data. As part of the results of such pilot study, the IDIS Small Bodies and Dust node developed a search system based on a preliminary XML data model. Here we introduce the goals of the IDIS initiative and describe the structure and the working of this search system. The source code of the search system is released under GPL license to allow people interested in participating to the IDIS initiative both as developers and as data providers to familiarise with the search environment and to allow the creation of volunteer nodes to be integrated into the existing network.

## Introduction

Recent years have been extremely eventful in the field of planetary sciences, with several space missions devoted to the exploration of the Solar System producing large amounts of high-quality data. Moreover, ground-based and near Earth-based observations, both through stand-alone projects and in coordination with space missions, contributed to increase the information we possess on Solar System. Those data are extremely heterogeneous both in their physical contents (e.g. electromagnetic data, gravitational data, geometric data and state vectors, etc.) and in their representations (images, spectra, numerical data). While such wealth of data can allow us to get an integrated picture of the origin and evolution of planetary systems, it should be taken into account that these data are spread over several specialised communities and archived through several different formats. The lack of interoperability between the different formats and the spreading of the data act against their effective use: to counteract this effect, several initiatives were started inside the scientific community involved in the study of the Solar System. One of such initiatives is the *Planetary Data System (PDS)* sponsored by NASA (http://pds.jpl.nasa.gov/). PDS is a distributed archive of Solar System data, with a focus on those data produced by space missions, prepared in a standard format designed to ensure the long-term access to the information. Another initiative is that of the *Space Physics Archive Search and Extract (SPASE) Consortium* (http://www.spase-group.org/index.jsp), supported by the space and solar physics communities. SPASE is an XML-based data system designed to build distributed data systems on heterogeneous networks of platforms and systems.

In the framework of the Sixth Framework Programme (FP6) for Research and Technological Development of the European Community, the *Europlanet (EPN)* project (http://europlanet.cesr.fr/) started the *Integrated and Distributed Information Service (IDIS)* initiative, also labelled as N7 activity. The goal of this initiative was to "...offer to the planetary science community a common and user-friendly access to the data and information produced by the various types of research activities: earth-based observations, space observations, modelling and theory, laboratory experiments. IDIS will also interface with existing databases and provide on-line products...." (from Europlanet N7 web-page, http://europlanet.cesr.fr/n7). The Europlanet project started in 2004 and ended its activities of 2008: during its lifetime, the FP6-funded IDIS initiative started a pilot study on how to create a XML-based distributed system of metadata portals for such heterogeneous and diffuse data. To achieve this goal, four scientific nodes, representative of a significant fraction of the scientific themes covered by planetary sciences, were created: the Interiors and Surfaces node, the Atmospheres node, the Plasma node and the Small Bodies and Dust node. They were complemented by a Technical node, with the aim of supporting and coordinating the activities of the scientific nodes concerning the design and the implementation strategy of the XML-based distributed system. In 2009, after the end of FP6 funding scheme, the original Europlanet program has been extended into the Europlanet Research Infrastructure (EPN-RI) project (http://www.europlanet-ri.eu/), funded by the Seventh Framework Programme (FP7) for Research and Technological Development. The IDIS initiative has been renewed under EPN-RI (http://www.europlanet-ri.eu/idis) and a new thematic node, the Planetary Dynamics node, was added. Starting from the results of EPN-N7 pilot study, the goal of the IDIS initiative under FP7 is to create a distributed system of metadata portals.

Here we describe the XML-based search system developed at the Small Bodies and Dust node (hereafter SBDN) as a result of the pilot study performed by the IDIS initiative during FP6. In Section 1 we will briefly describe the data management system over which we built the search system, in Section 2 we will detail the user interface to access the search system while in Section 3 we will describe the working of the different query facilities supplied by the search system. Since the IDIS initiative is strongly based on voluntary participation of data providers and since the existing thematic nodes cover most but not all the scientific fields covered by planetary sciences, in concluding our presentation we will point the readers interested in contributing to the IDIS efforts to the location where to download the source code of our search system.

## Section 1: XML-based data management system

The database management system adopted by IDIS-SBDN search system is eXist (http://exist-db.org). eXist is an open source database management system built on XML technology and distributed under GNU LGPL license. eXist supports different web standards, e.g. XQuery 1.0, XPath 2.0, XSLT 2.0, REST, SOAP, etc. and provides a development environment for web applications based on XQuery. The version over which we built the IDIS-SBDN search system is the 1.2.3 release. Here we will limit ourselves to a brief review of the features used to build IDIS-SBDN search system: for further details on eXist and its working we refer the readers to the documentation available at the homepage of the program, http://exist-db.org, or alternatively at http://exist.sourceforge.net.

### 1.1 eXist and XQuery

The IDIS-SBDN search system relies on the XQuery 1.0 technology, which in turn is based on the XPath 2.0 formalism. The core of eXist is the XQuery support, which is compliant to the W3C XML Query Working Group standards. XQuery allows to manage the XML database: it is designed to allow concise and easily understandable queries and to be sufficiently flexible to handle a broad spectrum of XML information sources, including both databases and documents.

The descriptions of the resources are stored using an XML tree structure: each element of the structure is called a *node* and is identified by a unique node identity. The knowledge of the whole tree structure is mandatory to access each single node.

The information is extracted from the node through the XPath language, which operates on the data model to provide the result of the search process. The basic building block of XPath is the expression, i.e. a string of characters: the language provides several ways to build up an expression starting from keywords, symbols and operands and it also allows nested expressions.

## 1.2 The Sandbox

eXist environment provides the developer with a web application called *Sandbox*, which allows to test the XQuery functions created to retrieve information from the database. While Sandbox has been supplied mainly as a template eXist application, we used it to debug and validate the xQuery functionalities of our search facilities (see Section 3). The testing was performed by sending queries to the local database previously loaded into the eXist archive. Once validated by Sandbox, the xQuery search subroutines were incorporated into the search system.

## 1.3 Data organisation

Resources are stored in XML files: the data model over which the present implementation of the IDIS-SBDN search system is based is a temporary XML porting of release 1.3.2 of the "Resource Submission Form" developed by the N7 activity of the Europlanet-FP6 project and accessible at <http://europlanet.cesr.fr/n7/res/Resource/CreateUpdate/>. This temporary data model has been adopted while the official XML-based data model for planetary science is developed in the framework of the FP7-funded Europlanet-RI program.
A Java application, called *client*, is provided by eXist to manage resources stored within the database, e.g. adding/removing resources from a database or creating a new database. This simple client is used to manage databases by means of a file system viewer window which allows users to create/update/remove files and folders.
The data model consists of the following XML files:
- **Activity**: contains data about the activities of the thematic nodes of Europlanet;
- **Country**: the list of the world countries;
- **Institute**: the list of the institutes involved in the Europlanet project;
- **Keyword**: contains the names and the type-ID of the keywords in the descriptions of the resources;
- **KeywordType**: contains the values of the type-ID values present in Keyword;
- **N2dwg**: the list of the Discipline Working Groups (DWG) from the Europlanet N2 activity;
- **Node**: the list of the names and web addresses of the Europlanet nodes;
- **PDSnode**: the list of the names and web addresses of the PDSnodes;
- **Person**: the list of people involved in the Europlanet and Europlanet-RI project with related information;
- **Resource**: the list of the resources related to a Europlanet node;
- **ScienceCase**: the list of the science cases and their description.

Each XML file is composed by several nodes: these nodes contain the description of the resources and an identifier (ID) to connect related resources in different files. These IDs is index every resource into the database.

## Section 2: IDIS-SBDN search system

The IDIS-SBDN search system has been developed to catalogue and supply a user-friendly environment to browse the existing resources for the different fields of planetary sciences. The IDIS-SBDN system is based on *resources* XML descriptors, which at present are divided into two main sections: the *General Information* (see fig. 1) and *EPN resources* (see fig. 2). The first section refers to general information about Europlanet, e.g. involved institutes, performed activities, details on the nodes which constitute the IDIS system, activities of Europlanet groups. At present, this section does not offer any pre-defined values to be searched for: the user is required to input one or more values through a text-box.

(Fig.1 fig1.tif caption: View of the *General Information* section.)

The second category supplies information related to specific resources existent into Europlanet databases, e.g. persons, web sites, data-sets or databases. Inside this section, the search mask is divided in two sub-sections: the first one devoted to general search domains (i.e. "Person" and "Resource" in the present release) and the second one supplying a list of planetary sciences-related search domains. Such subdivision reflects the structure of the XML schema used to describe the resources.

(Fig.2 fig2.tif caption: View of the *EPN Resources* section. Search domains are divided into two main categories. The one which uses pre-defined keywords is shown.)

Due to its general nature, the search mask of the first subsection does not supply any pre-defined search key: like in the case of the "General information" section previously described, the user is supplied with a text-box to enter the search values. The search mask of the second sub-section, instead, supplies the user with a pre-determined list of search values varying with the selected search domain (i.e. the kind of resources the user is interested in).

## 2.1 On-line help & suggestion service

The constant striving to ensure the user-friendliness of web sites and web portals introduced as a de facto standard the inclusion of on-line, self-explanatory suggestion and help services to allow the users to fully take advantage of the offered services and to make the learning curve smoother.
In order to simplify its use in the creation of user-defined queries, the IDIS-SBDN search engine activates a suggestion service every time one of the text-box based search domains is selected. When the user starts to enter the search key into the text-box, the system searches within the node database for all words starting with the entered characters and shows them to the user as a list which appears under the text-box itself. This is achieved by sending to the node server a series of AJAX requests while holding the text-box content: the system then executes the relevant xqueries (see 3.2) to look for the typed values within predefined fields.
In addition to this suggestion service, the IDIS-SBDN search has been complemented with an exhaustive on-line help (see fig. 3) which supplies the users with a description of the structure of the system, the procedure to create queries and the meaning of the different sections and fields.

(Fig.3 fig3.tif caption: View of the on-line help. Main menu is present on the left side for arguments selection.)

## 2.2 The search results

The idea at the basis of IDIS system is the creation of a network of thematic nodes, each with its own independent database of resources, and a common search protocol which would allow to merge them into a network-wide search system.
This idea has been considered as central during the development phase of the result page of the IDIS-SBDN search engine, which has been designed to enhance the presentation of the results. In order to assure the relevancy of the results to the submitted queries, the system presents separately those retrieved on the local node (i.e. those most likely of interest to the user) from the one retrieved from remote nodes, possibly of interest but related to different scientific themes than the one selected by the user.

### 2.2.1 Results layout

A central point during the development of the IDIS-SBDN search system was the design of the page layout. The page has been divided into two parts. The first one contains the results retrieved from local node and it dominates the page to emphasise its pertinence to the actual query. The second one contains the results coming from remote nodes and is presented as a secondary element in the page.

A second point to be accounted for was the amount of information presented for both types of results. We adopted the following policy: all the available information is showed for each local result while for remote results the system visualise only the number of resources found in the databases of the remote nodes, without showing any other information about them.
This policy has the goal of stressing the difference between the two kinds of results and of making immediately visible the presence of a primary and of a secondary element in the result page. At the top of the page the global title reports the main elements characterising the query: the selected resource domain and the inserted values. An example of the page layout and how the results are presented is shown in Fig.4.

(Fig.4 fig4.tif caption: Example of the results page. Results from local node are placed on the left side of the window. Results from other nodes are placed on the right side of the window.)

## 2.2.2 Local results

The *Local Query Facility* (LQF in the following) reports to users the resources existing at the local node which satisfy the user's query parameters: the results are presented through a card system resuming all the available information. Each card is topped by a title bar reporting the basic information of the resource: its name, its brief description and the URL of the related web page, if existing. The remaining information is divided into thematic sections accessible through dedicated tabs. This presentation of the data aims to allow the users to quickly check the results to verify their relevance and to easily browse the information. LQF is enhanced through a *Secondary Query Facility* (SQF in the following). Where possible, the fields describing each resource are complemented with a button depicting a question mark symbol: when the user clicks on such a button, the system automatically starts an AJAX query on the content of the field (see section 3.3). Information retrieved through this query is shown in a secondary window that appears next to the button. The card-based presentation of the results and the SQF has been designed to supply the users with a hierarchical search tool that would allow to efficiently explore the resources existing in the repository and their inter-correlations.

## 2.2.3 Remote results

The *Remote Query Facility* (RQF in the following) is part of the core of the IDIS infrastructure as it interconnects each node to all the others. RQF presently consists of some sections, one for each node composing IDIS. Each sections reports the number of results obtained by a node through a remote query performed by the local node: to ease the comparisons between the number of results at the different nodes, also the number of those found at the local node is shown. Each section contains a link to the homepage of the relevant node and, if the number of results is greater than zero, the number itself is a link to the page of the remote query results.

# Section 3: IDIS-SBDN search engine and query facilities

The general mechanism of the IDIS-SBDN search system is quite simple and is divided in two components: the LQF and the RQF.
Both facilities work in the same way through a client-server approach: information (the query) is sent to a server (independently of it being local or remote) using the *keyword=value* format and then the answer (the results of the query) is collected by the client in XML or HTML format. The information sent to the server contains the selected resource type, the inserted values and the query type (LQF, RQF or SQF)

## 3.1 The Local Query Facility

The Local Query Facility sends a single query to the local node to retrieve all available resources (and the related information) that satisfy the user's query. All keyword/value pairs inserted at the search page are used to create the appropriate query.
The main search function handles the different types of resources to use the appropriate query for each of them. The first important distinction is between the *EPN Resources* and the *General Information* classes, previously mentioned in section 2 when discussing the structure of the search page, due to the different complexity of their XML descriptions. *General Information* resources have a simpler XML structure while *EPN Resources* are characterised by a more complex one possessing child nodes.
Queries related to *EPN Resources* are divided into two main classes. The first class concerns queries related to *Person* or *Resource* objects while the second one collects all the queries related to the other resources types, this division being due to the different insertion tools employed. For the first class of resources, the LQF supplies a text-box in the search page to allow the user to input one or more search keys. For the second class, the LQF supplies a list of predefined values once the user selects the desired resource kind. Consequently, in those queries belonging to the second class only one XML field, corresponding to the value selected by the user (e.g., mission, scientific field), is checked while in queries belonging to the first class all of the XML fields are checked.
For the searches related to the *General Information* class, the query mechanism is analogous to the one already described for the *Person* and *Resource* objects.
The resources retrieved as the result of a query are packed into an XML file containing both the query keyword/value couples and query results: this XML file is then transformed into an HTML page using the apposite XSLT converter.

## 3.2 The Remote Query Facility

The basic working of RQF is similar to that of LQF. The Remote Query Facility, however, directs the query to the remote nodes composing the IDIS infrastructure and retrieves information about the resources possibly satisfying the search parameters. The redirection of the query is performed through a native function supplied by the *eXist* database.
Unlike LQF, RQF shows only the number of found resources, not the actual list of resources. Consequently, no conversion into HTML is done from the XML file obtained by the remote servers. The retrieved XML file is read by a Javascript function which extracts the information and shows it through the relevant sections in the search web-page.

## 3.3 The suggestion query

To help the users in creating their queries, a suggestion service has been implemented in the IDIS-SBDN system.
As we mentioned in section 2.1, this service is activated each time a user selects a resource kind that requires the insertion of one or more values using the text-box. When any text is inserted in the text-box, the system dispatch it to the server of the local node through an AJAX query. The contacted server then performs an XQuery looking for possible suggestions. Two are the information sent to the server: the selected kind of resource and the inserted text. The system then looks for all resources corresponding to the selected type that also contain the inserted text. To limit the workload and thus the possibilities of slowdowns, the system looks for suggestions within selected descriptors of the resources, i.e. name and description. This implies that, even if no suggestion is found, relevant resources containing the inserted text could anyway be present in the database.

## 3.4 The Secondary Query Facility

As already mentioned in section 2.2.2, to help the users in retrieving all information possibly related to their searches, the IDIS-SBDN system implements a Secondary Query Facility, which is accessible through the result page of each search (see fig. 5). The working of SQF is similar to that of LQF and employs the same functions: it is based on specific queries related to resources identified by a keyword/value couple corresponding to the resource name. The main difference between LQF and SQF is the use of the value of the resource ID instead of generic values to be found within the XML resource descriptor: such search strategy allows for a more direct localisation of the resources and assures better performances in the data retrieval.

(Fig.5 fig5.tif caption: Same results page as fig.4. Here the secondary query facility has been used to retrieve information about Rosetta mission.)

## Conclusions

In this paper we presented the search system developed at SBDN as part of the pilot study performed by the IDIS initiative during FP6. As we mentioned in the introduction, the goal of the study was to assess the issues concerning the design and implementation of a distributed search system interconnecting data repositories managed by different communities and characterised by different archiving strategies and different physical contents. We refer the readers interested in the development and the accomplishment of the IDIS initiative over both EPN and EPN-RI programs to the relevant web-portals, respectively http://europlanet.cesr.fr/ and http://www.europlanet-ri.eu/. Here we limited ourselves in describing the tools developed by SBDN: the source code of the search system and its user interface are released under GPL 3.0 license and are freely available at SBDN website, http://www.ifsi-roma.inaf.it/europlanet/. These tools, like most products of IDIS initiative under EPN program, represent only a very preliminary version of the ones that will compose the infrastructure to be developed under EPN-RI. They are publicly released with the main goals to show to the community what the IDIS initiative aims to accomplish and to attract data suppliers and scientists willing to volunteer their participation to the IDIS project itself.

## Acknowledgements

The development of the IDIS-SBDN search system has been performed in the framework of the N7 IDIS activity of the European Planetology Network (EuroPlaNet) project. The European Planetology Network project has been supported by the European Community in the Sixth Framework Programme (FP6) for Research and Technological Development through contract n. 001637. The Europlanet Research Infrastructure project is supported by the European Community in the Seventh Framework Programme (FP7) for Research and Technological Development through grant agreement n. 228319. The IDIS Small Bodies and Dust thematic node (SBDN) is hosted at the Institute for Physics of Interplanetary Space (IFSI) in Rome and jointly operated with the Institute for Astrophysics and Cosmic Physics (IASF) in Rome. Both IFSI and IASF are part of the Italian National Astrophysics Institute (INAF). The development and hosting of the IDIS-SBDN search system have been partially financed by INAF. The authors wish to thank M. T. Capria for useful discussions and valuable suggestions to improve the paper.

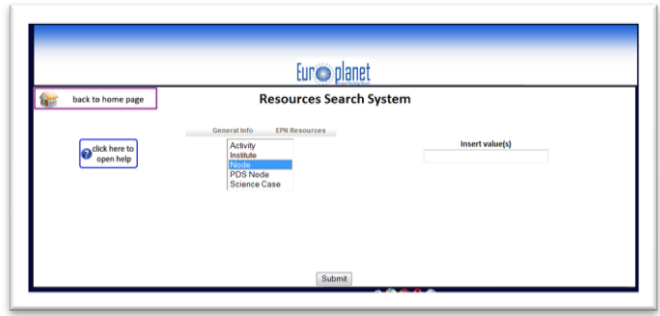

fig.1

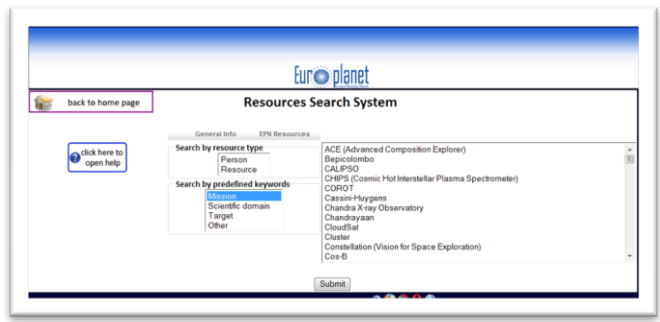

fig.2

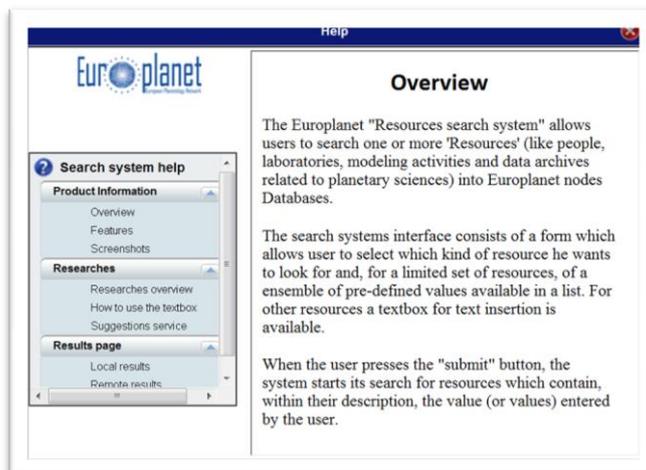

fig.3

fig.4

fig.5